\documentclass{ws-ijmpe}

\begin{document}

%\markboth{Authors' Names}{Instructions for  
%Typing Manuscripts (Paper's Title)}

\catchline{}{}{}{}{}
%%%%%%%%%%%%%%%%%%%%%%%%%%%%%%%%%%%%%%%%%%%%%%%%%%%%%%%%%%%%%

\title{Pion-nucleon scattering within a gauged linear sigma model with
parity-doubled nucleons}
\author{Susanna Wilms, Francesco Giacosa, and Dirk H.\ Rischke}
\address{Institut f\"ur Theoretische Physik\\  
Johann Wolfgang Goethe-Universit\"at\\ 
Max-von-Laue-Str.\ 1, D-60438 Frankfurt am Main, Germany}
\maketitle

\begin{history}
\received{(received date)}
\revised{(revised date)}
%\accepted{(Day Month Year)}
%\comby{(xxxxxxxxxx)}
\end{history}

\begin{abstract}
We compute pion-nucleon scattering at tree-level within a gauged
linear sigma model which contains the nucleon and its chiral partner.
Such an investigation in principle allows to make definite predictions
as to whether the main contribution to the nucleon mass comes from the
chiral condensate or from the mixing with its chiral partner.
We find that there seems to be no set of model parameters 
that allows for a simultaneous description of 
all experimentally measured scattering lengths and range parameters.
This indicates the need to improve the dynamical
ingredients of the model.
\end{abstract}

\section{Introduction}

Effective models which embody chiral symmetry and its
spontaneous breakdown at low temperatures and densities, are
widely used to understand the properties of light hadrons. 
Viable candidates obey a well-defined set of low-energy 
theorems \cite{meissner}, but they still differ in some 
crucial and interesting aspects such as the mass generation
of the nucleon and the behavior at non-zero $T$ and $\mu$.

Here we concentrate on a gauged linear sigma
model with $U(2)_R \times U(2)_L$ symmetry and parity-doubled nucleons. 
The mesonic sector involves scalar,
pseudoscalar, vector and axialvector mesons; Vector Meson Dominance
(VMD) automatically follows from gauging the symmetry group 
\cite{Ko,pisarski}. 
In the baryonic sector, besides the usual nucleon doublet field $N,$ a
second baryon doublet $N^{\ast }$ with $J^{P}=\frac{1}{2}^{-}$ is
included. As first discussed in Ref.\ \cite{DeTar:1988kn} and extensively
analyzed in Ref.\ \cite{jido}, in the so-called mirror 
assignment the nucleon fields $N$ and $N^{\ast }$ do not 
have vanishing mass in the chirally restored phase. The 
chiral condensate $\varphi $ further increases the nucleon masses 
and generates a mass splitting of $N$ and $N^{\ast },$ but is no
longer solely responsible for generating the nucleon masses
(see section 2 for details). Such a
theoretical set-up has been used in Ref.\ \cite{zschiesche} to study the
properties of cold and dense nuclear matter. The experimental assignment for
the partner of the nucleon is still controversial: the well-identified
resonance $N^{\ast }(1535)$ is one candidate, but a very broad and not yet
discovered resonance centered at about $1.2$ GeV has been proposed in Ref.\
\cite{zschiesche}.

The aim of the present work is the calculation of pion-nucleon scattering
at tree-level within the outlined model. This study constitutes a decisive test
for its viability and can provide useful information to clarify the origin
of the nucleon mass and its behavior in the chirally restored phase.
Furthermore, important issues such as the dependence of the scattering
lengths on the mass of the chiral partner $N^{\ast }$ and on the enigmatic $%
\sigma $ meson can be addressed.

This paper is organized as follows: in section 2 we briefly describe
the details of the model, in section 3 we present
and discuss the scattering amplitudes and finally, in
section 4, we summarize the present stage of our research and outline
future developments.

\section{The gauged linear sigma model with parity-doubled nucleons}

The scalar and pseudoscalar fields are included in the matrix $\Phi =(\sigma
+i\eta )t^{0}+(\overrightarrow{a}_{0}+i\overrightarrow{\pi })
\cdot \overrightarrow{t}$ and the vector and axialvector fields 
are represented by the matrices $R^{\mu}=(\omega ^{\mu }-f_{1}^{\mu })
t^{0}+(\overrightarrow{\rho }^{\mu }-
\overrightarrow{a_{1}}^{\mu })\cdot \overrightarrow{t}$ and $L^{\mu }=(\omega
^{\mu }+f_{1}^{\mu })t^{0}+(\overrightarrow{\rho }^{\mu }+\overrightarrow{
a_{1}}^{\mu })\cdot \overrightarrow{t}$ 
($\overrightarrow{t}=\frac{1}{2}\overrightarrow{\tau },$ where $%
\overrightarrow{\tau }$ are the Pauli matrices and $t^{0}=\frac{1}{2}1_{2}$).

The mesonic Lagrangian, as presented in Refs.\
\cite{Ko,pisarski}, is constructed by requiring 
invariance under local chiral transformations; this local
invariance is then broken to a global one by
mass terms for the vector and axialvector mesons:
\begin{eqnarray}
{\cal L}_{\text{mes}} &=&{\rm Tr}
\left[ (D_{\mu }\Phi )^{\dagger}(D^{\mu }\Phi )-m^{2}\Phi^{\dagger}\Phi 
-\lambda _{2}\left( \Phi ^{\dagger}\Phi \right) ^{2}\right] -
\lambda_{1}\left({\rm Tr}[\Phi ^{\dagger}\Phi ]\right)^{2}  \notag \\
&&+c\, (\det \Phi ^{\dagger}+\det \Phi )
+{\rm Tr}[H(\Phi ^{\dagger}+\Phi )]  \notag \\
&&-\frac{1}{4}{\rm Tr}\left[(L^{\mu \nu })^{2}+(R^{\mu \nu })^{2}\right]
+\frac{m_{\rho }^{2}}{%
2}{\rm Tr}\left[(L^{\mu })^{2}+(R^{\mu })^{2}\right]\; ,  \label{Lmes}
\end{eqnarray}%
where $D^{\mu }\Phi =\partial ^{\mu }\Phi +ig(\Phi R^{\mu }-L^{\mu }\Phi ).$

The baryon sector involves the baryon doublets $\Psi _{1}$ and $\Psi _{2},$
where $\Psi _{1}$ has positive parity and $\Psi _{2}$ negative parity. 
While the former transforms in the standard fashion under chiral 
transformations, the
latter is postulated to transform according to the mirror assignment, namely:
\begin{eqnarray}
\Psi _{1R} &\longrightarrow &U_R \Psi _{1R},\text{ }\overline{\Psi }%
_{1R}\longrightarrow \overline{\Psi }_{1R}U_R^{\dagger }\; ,  \notag \\
\Psi _{2R} &\longrightarrow &U_L\Psi _{2R},\text{ }\overline{\Psi }%
_{2R}\longrightarrow \overline{\Psi }_{2R}U_L^{\dagger }\; ,  \label{mirror}
\end{eqnarray}%
and similarly for the left-handed fields.
Such field transformations allow to write down the following chirally
invariant Lagrangian:
\begin{eqnarray}
{\cal L}_{\text{nucl}} &=&\overline{\Psi }_{1L}
i\gamma_{\mu }D^{\mu }_L \Psi
_{1L}+\overline{\Psi }_{1R}i\gamma_{\mu }D^{\mu }_R \Psi _{1R}+
\overline{\Psi }_{2L}i\gamma_{\mu }D^{\mu }_R \Psi _{2L}
+\overline{\Psi }_{2R}i\gamma_{\mu }D^{\mu }_L\Psi _{2R}  \notag \\
&&-\widehat{g}_{1}\left( \overline{\Psi }_{1L}\Phi \Psi _{1R}\ +\overline{%
\Psi }_{1R}\Phi ^{\dagger}\Psi _{1L}\right) 
-\widehat{g}_{2}\left( \overline{\Psi }_{2L}\Phi ^{\dagger}\Psi _{2R}\ 
+\overline{\Psi }_{2R}\Phi \Psi _{2L}\right) 
\notag \\
&&-m_{0}(\overline{\Psi }_{1L}\Psi _{2R}-\overline{\Psi }_{1R}\Psi _{2L}-%
\overline{\Psi }_{2L}\Psi _{1R}+\overline{\Psi }_{2R}\Psi _{1L})\; ,
\label{lnuc}
\end{eqnarray}
where $D^\mu_R = \partial^\mu - i g R^\mu$,
$D^\mu_L = \partial^\mu - i g L^\mu$.

The last term, whose strength is parameterized by $m_{0},$ plays a crucial
role in the model. It is a quadratic mixing term for the nucleon fields 
$\Psi _{1}$ and $\Psi _{2}.$ The physical fields $N$ and $N^{\ast },$
referring to the nucleon and to its chiral partner, arise by
diagonalizing the Lagrangian ${\cal L}_{\text{nucl}}$ 
and are a superposition of
the two fields $\Psi _{1}$ and $\Psi _{2}$:
\begin{equation}
\left( 
\begin{array}{c}
N \\ 
N^{\ast }
\end{array}
\right) =\frac{1}{\sqrt{2\cosh \delta}}\left( 
\begin{array}{cc}
e^{\delta /2} & \gamma _{5}e^{-\delta /2} \\ 
\gamma _{5}e^{-\delta /2} & -e^{\delta /2}%
\end{array}
\right) \left( 
\begin{array}{c}
\Psi _{1} \\ 
\Psi _{2}
\end{array}
\right) .  \label{mixing}
\end{equation}
The parameter $\delta ,$ which measures the intensity of the mixing, is
expressed in terms of the parameters of ${\cal L}_{\text{nucl}}$ as:%
\begin{equation}
\sinh \delta =\frac{(\widehat{g}_{1}+\widehat{g}_{2})\varphi }{4m_{0}}\; ,
\label{delta}
\end{equation}%
where the coupling constants $\widehat{g}_{1}$ and $\widehat{g}_{2}$
parametrize the interaction of the baryonic fields with scalar and
pseudoscalar mesons and $\varphi =$ $\left\langle 0\left\vert \sigma
\right\vert 0\right\rangle =Z f_{\pi }$ is the chiral condensate.
The parameter $f_{\pi }=92.4$ MeV is
the pion decay constant and $Z=m_{a_{1}}/m_{\rho }\approx 1.59.$ The latter
arises by shifting the isotriplet axialvector meson field $\overrightarrow{a%
}_{1}^{\mu }$ in order to avoid an unphysical mixing with the pion 
\cite{gasioro}.

The masses of the physical fields $N$ and $N^{\ast }$ read: 
\begin{equation}
m_{N,N^{\ast }}=\displaystyle\sqrt{m_0^2 + \left[\frac{(\widehat{g}_{1}+
\widehat{g}_{2})\varphi}{4}\right]^2} \pm 
\frac{(\widehat{g}_{1}-\widehat{g}_{2})\varphi }{4}\; .  \label{masses}
\end{equation}%
In Eq.\ (\ref{masses}) both $m_{0}$ and $\varphi $ contribute to the masses
of the nucleon and its partner. Some considerations about two important
limiting cases of the above listed equations are in order:

(i) When $\delta \rightarrow \infty $, corresponding to 
$m_{0}\rightarrow 0,$ no mixing is present and $N=\Psi _{1},$ $N^{\ast }=\Psi
_{2}.$ In this case $m_{N}=\widehat{g}_{1}\varphi /2$ and $m_{N^{\ast }}=%
\widehat{g}_{2}\varphi /2$, thus the nucleon mass is generated solely by the
chiral condensate as in the linear sigma model.

(ii) In the chirally restored phase where $\varphi \rightarrow 0$, 
one has mass degeneracy $m_{N}=m_{N^{\ast }}=m_{0}.$ 
When chiral symmetry is broken, $\varphi \neq 0$, a splitting is
generated. By choosing $0<$ $\widehat{g}_{1}<\widehat{g}_{2}$ the inequality 
$m_{N}<m_{N^{\ast }}$ is fulfilled.

In Ref.\ \cite{zschiesche} a large value of the parameter $m_{0}$ ($\sim 800$
MeV) is claimed to be needed for a correct description of 
nuclear matter properties, thus
pointing to a small contribution of the chiral condensate to the nucleon
mass. Validating this claim through the evaluation of pion-nucleon
scattering at zero temperature and density is subject of the present paper.

\section{The scattering amplitudes}

The general form of the Lorentz-invariant scattering amplitude can be
written as \cite{matsui}:
\begin{equation}
T_{ab}=[A^{(+)}+\frac{1}{2}(q_{1}^{\mu }+q_{2}^{\mu })\gamma _{\mu
}B^{(+)}]\delta _{ab}+[A^{(-)}+\frac{1}{2}(q_{1}^{\mu }+q_{2}^{\mu })\gamma
_{\mu }B^{(-)}]i\epsilon _{bac}\tau _{c}\; ,  \label{Tab}
\end{equation}%
where the subscripts $a$ and $b$ refer to the isospin of the initial and
final states and the superscripts $(+)$ and $(-)$ denote the isospin-even
and isospin-odd amplitudes, respectively.

The pion-nucleon scattering amplitudes, evaluated from 
the Lagrangian ${\cal L}_{\text{nucl}}$ of Eq.\ (\ref{lnuc}) at tree-level,
involve exchange of $\sigma $ and $\rho$ mesons in the $t$-channel 
and intermediate $N$ and $N^{\ast }$ states in the $s$- and $u$-channels. 
At threshold, $t=0,$ $s=(m_{N}+m_{\pi })^{2},u=(m_{N}-m_{\pi })^{2}$,
the explicit expressions for the terms $A^{(\pm )}$ and $B^{(\pm )}$ 
read ($r=m_{N^{\ast }}/m_{N}$, $y = m_\pi/m_N$, threshold values are
indicated by a subscript ``0''):
\begin{eqnarray}
A_{0}^{(+)} &=&\frac{m_{N}}{Z^{4}f_{\pi }^{2}}
\left\{\frac{(r-1)(r^{2}-1)}{2\cosh^{2}\delta}
\frac{r^{2}-1-y^2}{(r^{2}-1-y^2)^{2}-4y^2} \right. \nonumber \\
&&+\left. \left( 1-\frac{r+1}{2\cosh ^{2}\delta}\right) 
\left[ 1+\left( \frac{m_{\pi }}{m_{\sigma }}\right)^{2}(Z^{2}-2)\right] 
\right\}\; , \label{A0+} \\
A_{0}^{(-)}& =& \frac{m_{\pi }}{Z^{4}f_{\pi }^{2}\cosh^{2}\delta}
\frac{(r-1)(r^{2}-1)}{(r^{2}-1-y^2)^{2}-4y^2}\; , \\
B_{0}^{(+)}& =& \frac{m_{N}}{m_{\pi }Z^{4}f_{\pi }^{2}}
\left[\frac{(r-1)^{2}}{\cosh^{2}\delta}
\frac{y^2}{(r^{2}-1-y^2)^{2}-4y^2}
-\frac{\tanh ^{2}\delta}{1-y^2/4}\right]\; ,\\
B_{0}^{(-)}& =& \frac{1}{2Z^{4}f_{\pi }^{2}}
\left[ Z^{4}-1+\frac{\tanh ^{2}\delta}{1-y^2/4}
+\frac{(r-1)^{2}}{\cosh^{2}\delta}
\frac{r^{2}-1-y^2}{(r^{2}-1-y^2)^{2}-4y^2}\right]\; .
\end{eqnarray}
The $s$-wave and $p$-wave scattering lengths,
$a_{0}^{(\pm )}$ and $a_{1^\pm}^{(\pm )} $, and the 
effective $s$-wave range of the interaction, $r_0^{(\pm)}$, are given by:
\begin{eqnarray}
a_{0}^{(\pm )}&=& \eta \left( A_{0}^{(\pm )}+m_{\pi }B_{0}^{(\pm )}
\right)\; ,\label{a0} \\
a_{1^{+}}^{(\pm )}& =& \frac{2}{3}\eta\, C_{0}^{(\pm )}\; , \\
a_{1^{-}}^{(\pm )}&= & \frac{2}{3}\eta \, C_{0}^{(\pm )}
-\frac{\eta }{4m_N^{2}} \left[A_{0}^{(\pm )}-
(2m_{N}+m_{\pi })B_{0}^{(\pm )} \right]\; ,  \label{a1+} \\
r_{0}^{(\pm )} &=&\eta \left\{ -2\, 
C_{0}^{(\pm )}+\frac{(m_{N}+m_{\pi })^{2}}{%
m_{N}m_{\pi }}D_{0}^{(\pm )} \right. \notag \\
&&-\left. \frac{1}{2m_{N}m_{\pi }}\left[ \left(1-\frac{m_{\pi }}{2m_{N}}\right)
A_{0}^{(\pm)}-\left(m_{N}+\frac{m_{\pi }^{2}}{2m_{N}}\right)
B_{0}^{(\pm )}\right] 
\right\}\; , \label{range}
\end{eqnarray}
where $\eta =1/[4\pi (1+y)]$,
$C_{0}^{(\pm )}= 
\frac{\partial }{\partial t}(A^{(\pm )}+m_{\pi }B^{(\pm )})|_{t=0}$, and
$D_{0}^{(\pm )}=\frac{\partial }{\partial s}
(A^{(\pm )}+m_{\pi}B^{(\pm )})|_{t=0}$.
Note that the $s$-wave scattering lengths $a_{0}^{(\pm )}$ 
correctly vanish in the chiral limit, $m_\pi 
\rightarrow 0$. 

In a first study, we identify $N^{\ast }$ with the resonance $N^{\ast }(1535)$
and we fix the parameter $Z=m_{a_{1}}/m_{\rho }\approx 1.59.$ The scattering
lengths and the range parameter are shown 
as function of the mixing parameter $\delta $ in Fig.\ 1 
for different values of the $\sigma $ mass. 
In the isospin-even channel, 
we observe a dependence on $m_\sigma$,
which is particularly strong for
$a_{0}^{(+)}$ and $r_0^{(+)}$. 
However, when lowering the parameter $Z$ to $Z=\sqrt{2},$ as
predicted by the KSFR relation \cite{Ksfr}, the dependence on
$m_\sigma$ vanishes. This can be seen directly from Eq.\ (\ref{A0+}).
The quantities for the isospin-odd channel
are always independent of the value of
the $\sigma$ mass.

%%%%%%%%%%%%%%%%%%%%%%%%%%%%%%%%%%%%%%%%%%%%%%%%%%%%%%%%%%
\begin{figure}[th]
\centerline{\psfig{file=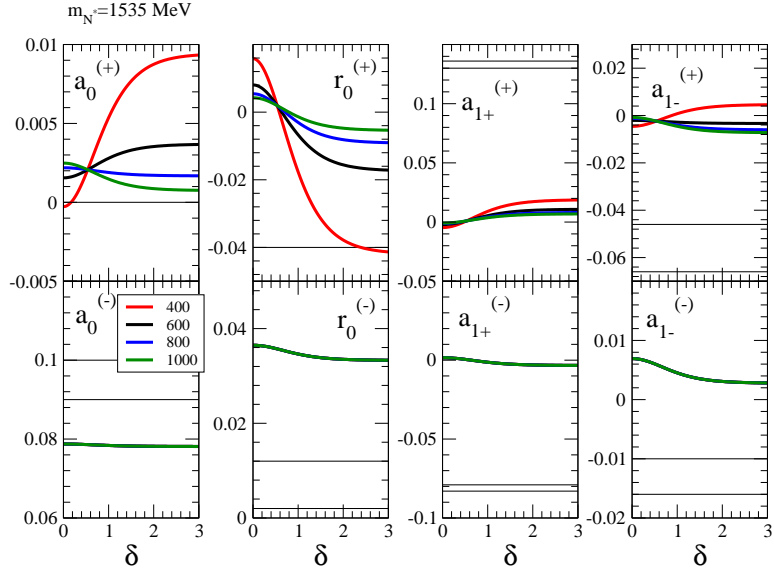,width=9cm,angle=270}} 
\vspace*{5pt}
\caption{Scattering lengths and range parameters
as function of $\protect\delta $, for different
values of $m_{\protect\sigma }$. The horizontal lines represent
the $1\sigma$ boundaries of the experimentally measured values.}
\end{figure}
%%%%%%%%%%%%%%%%%%%%%%%%%%%%%%%%%%%%%%%%%%%%%%%%%%%%%%%%%%%%%%%%%%%%%%%%%%%%%

A first, preliminary comparison of our results
to the experimental data on $\pi N$
scattering lengths, as measured in Ref.\ \cite{schroder} by precision X-ray
experiments on pionic hydrogen and pionic deuterium, yields
the following picture. (i) For small $m_\sigma\approx 400- 600$ MeV, 
a small value
of $\delta$ is favored, in order for $a_0^{(+)}$ to be close
to the experimental data. In this case, the dominant contribution to
the nucleon masses comes from $m_0$. (ii) For large $m_\sigma$, 
a large value for $\delta$ is required such that $a_0^{(+)}$ is
closer to the experimental data. This points towards 
a dominant contribution of the chiral condensate to the nucleon
mass. (iii) The effective range parameters
are closer to the experimental data for large $\delta$; here,
a small value for $m_\sigma$ is clearly favored.
(iv) All $p$-wave scattering lengths come out too
small in magnitude in order to compare reasonably well to 
the experimental data. This indicates the need to include
other baryonic resonances, such as the $\Delta$.

In summary, the comparison of characteristic quantities for
pion-nucleon scattering
to experimental data does not unambiguously favor definite values
for $m_\sigma$ and $\delta$.
Thus, it is not (yet) possible to make a clear-cut prediction
as to whether the mass of the nucleon is dominantly generated by
the chiral condensate or by mixing with its chiral partner.

\section{Summary and outlook}

In this paper we have computed the pion-nucleon scattering lengths at 
tree-level in the framework of a gauged linear sigma model 
with parity-doubled nucleons. Within
the mirror assignment the mass of the nucleon originates only partially
from the chiral condensate and does not vanish in the chirally restored
phase.
Quantitative predictions to test this scenario represent an important topic
in modern hadron physics. 

In a first study, we were not able to identify a
definite set of parameters of our model that would simultaneously
describe all scattering data. Therefore, a 
careful analysis of the dependence of the results on three
key parameters is in order: the parameter $Z,$ equal to $m_{a_{1}}/m_{\rho }$
in the mesonic sector of the Lagrangian but predicted to be $\sqrt{2}$ by
the KSFR relation, the mass of the $\sigma$ meson, $m_{\sigma },$ which cannot
be precisely determined due to the broadness of the 
corresponding resonance $f_{0}(400$-$1200)$, 
and the mass of the chiral partner of the nucleon 
$N^{\ast }$, which has not been yet unambiguously identified. Indeed, useful 
information toward its identification can be extracted. 
Then, by comparing the theoretical predictions with 
the experimental results we plan
to determine, within our model, to which degree the nucleon mass 
originates from the chiral condensate and how this point is
affected by the choice of the above mentioned parameters.

In a second step, we plan to extend our model by 
including the 
Weinberg-Tomozawa term \cite{wt} which is necessary to correctly 
reproduce the axial coupling
constant $g_{A}$.
Further applications, such as the inclusion of the $\Delta$ resonance 
and the study of radiative reactions, as the 
$\eta$-photoproduction, are also planned.

\section*{Acknowledgements}

The authors thank T.\ Kunihiro, J.\ Schaffner-Bielich, S.\ Str\"uber,
and D.\ Zschiesche for useful discussions.

\end{document}